\documentstyle[aas2pp4]{article}
\batchmode
\begin{document}
\def \nh {$N_{\rm H}$}
\def \nhgal {$N_{\rm H~I}^{\rm {\tiny Gal}}$}
\def \hb {H$\beta$}
\title{ THE SOFT X-RAY PROPERTIES OF A COMPLETE SAMPLE OF 
OPTICALLY SELECTED QUASARS\\ II. FINAL RESULTS}
\author {Ari Laor}
\affil{Theoretical Astrophysics,Caltech 130-33, Pasadena, CA 91125\\
Current address: Physics Department, 
Technion, Haifa 32000, Israel\\
laor@physics.technion.ac.il}
\author{Fabrizio Fiore}
\affil{Osservatorio Astronomico di Roma,
via dell'Osservatorio 5, Monteporzio-Catone (RM), 00040 and 
SAX Scientific Data Center, via corcolle  19 I-00131 
Roma, Italy\\
fiore@susy.mporzio.astro.it}
\author{Martin Elvis, Belinda J. Wilkes \and Jonathan C. McDowell}
\affil{Harvard-Smithsonian Center for Astrophysics, 60 Garden Street, 
Cambridge, MA 02138\\
elvis, belinda, mcdowell@cfa.harvard.edu }

\begin{abstract}
We present the final results of a {\em ROSAT} PSPC program
to study the soft X-ray emission properties of a complete sample of
of low $z$ quasars. This sample includes all 23 quasars from
the Bright Quasar Survey with  $z\le 0.400$, 
and \nhgal$< 1.9\times 10^{20}$~cm$^{-2}$. Pointed {\em ROSAT} PSPC 
observations were made for all quasars, yielding high S/N spectra
for most objects which allowed an accurate determination of the
spectral shape. The following main results were obtained:

\begin{enumerate}
\item The spectra of 22 of the 23 quasars are consistent, to within
$\sim 30$\%, with a single power-law model at rest-frame $0.2-2$~keV. 
There is no evidence for significant soft excess emission with 
respect to the best fit power-law. We place a limit (95\% 
confidence) of 
$\sim 5\times 10^{19}$~cm$^{-2}$ on the amount of excess foreground 
absorption by cold gas for most of our quasars. The limits are
$\sim 1\times 10^{19}$~cm$^{-2}$ in the two highest S/N spectra. 

\item The mean $0.2-2$~keV continuum of quasars agrees remarkably 
well with an extrapolation of the mean 1050\AA-350\AA\ continuum 
recently determined by Zheng et al. (1996) for $z>0.33$ quasars.
This suggests that there is no steep soft component below 0.2~keV.

\item Significant X-ray absorption ($\tau>0.3$) by partially ionized gas 
(``warm absorber'')
in quasars is rather rare, occurring for $\lesssim 5$\% of the population,
which is in sharp contrast to lower luminosity Active Galactic Nuclei 
(AGNs), where significant absorption
probably occurs for $\sim 50$\% of the population.

\item Extensive correlation analysis of the X-ray continuum emission 
parameters with optical emission line 
parameters indicates that the strongest correlation
is between the spectral slope $\alpha_x$, and the \hb\ FWHM. 
A possible explanation for
this remarkably strong correlation is a dependence of $\alpha_x$ on 
$L/L_{\rm Edd}$, as seen in Galactic black hole candidates.

\item The strong correlations between $\alpha_x$, and  
$L_{\rm [O~III]}$, Fe~II/\hb, and the peak [O~III] to \hb\ flux
ratio are verified. The physical origin of these correlations is
still not understood.

\item There appears to be a distinct class of ``X-ray weak'' quasars,
which form $\sim 10$\% of the population (3 out of 23),
where the X-ray emission is smaller by a factor of 10-30 than expected
based on their luminosity at other bands, and on their \hb\ luminosity.
These may be quasars where the direct X-ray source is obscured, and
only scattered X-rays are observed.

\item Thin accretion disk models cannot reproduce the observed
0.2-2~keV spectral shape, and they also cannot reproduce the tight
correlation between the optical and soft X-ray emission. An as yet
unknown physical mechanism must be maintaining a strong correlation
between the optical and soft X-ray emission.

\item The H~I/He~I ratio in the high Galactic latitude ISM must be
within 20\%, and possibly within 5\%, of the total H/He ratio of 10, 
which indicates
that He in the diffuse H~II gas component of the interstellar medium 
is mostly ionized to He~II or He~III.
\end{enumerate}

We finally note the intriguing possibility that although
$\langle\alpha_x\rangle$ in radio-loud quasars 
($-1.15\pm0.14$) is significantly 
flatter than in radio-quiet quasars ($-1.72\pm0.09$) the X-ray emission 
may not be related to the presence of radio 
emission. The difference in $\langle\alpha_x\rangle$ may result from the strong
$\alpha_x$ vs. \hb\ FWHM correlation and the tendency of radio-loud quasars
to have broader \hb.
\end{abstract}

\section{INTRODUCTION}
 
Quasars emit most of their power in the UV to soft X-ray regime.  
The PSPC detector aboard {\em ROSAT} allowed a significantly improved 
study of the soft X-ray emission of quasars compared with earlier missions
(some of which were not sensitive below 2~keV),
such as {\em HEAO-1, EINSTEIN, EXOSAT}, and {\em GINGA} 
(e.g. Mushotzky 1984; Wilkes \& Elvis 1987; Canizares \& White 1989; 
Comastri {\it et al.} 1992; Lawson {\it et al.} 1992; Williams {\it et al.} 
1992; and a recent review by Mushotzky, Done, \& Pounds 1993).
These earlier studies indicated that the X-ray emission above 1-2~keV 
is well described by a power law with a spectral slope 
$\alpha_x= d\ln f_{\nu}/d\ln \nu$ of about $-0.5$ for radio loud quasars and
about $-1.0$ for radio quiet quasars. Large heterogeneous samples of AGNs
were recently studied using the {\em ROSAT} PSPC by Walter \& Fink (1993) and
by Wang, Brinkmann, \& Bergeron (1996). However, the objects studied in 
the papers mentioned above do not form a complete sample, and the available 
results may be biased by various
selection effects which were not well defined a priori. In particular, most 
of the studied objects are nearby, intrinsically X-ray bright AGNs.

To overcome the potential biases in existing studies we initiated
a {\em ROSAT} PSPC program to make an accurate determination of the soft 
X-ray properties of a well defined and complete sample of quasars,
selected independent of their X-ray properties.
 This program was designed 
to address the following questions:
\begin{enumerate}
\item What are the soft X-ray spectral properties of  
the low redshift quasar population?
\item Are simple thin accretion disk models (e.g. Laor 1990) able to fit the
 observed 
optical/UV/soft X-ray continuum? are other modifying mechanisms, such as a 
hot corona (e.g. Czerny \& Elvis 1987) required?  Are models invoking 
optically thin free-free emission possible (e.g. Ferland, Korista \&
Peterson, 1990; Barvainis 1993)?
\item Do the observed soft X-ray properties display any significant correlations 
with other properties of these quasars? Are these correlations compatible with
various models for the continuum and line emission mechanisms?
\end{enumerate}

Our sample includes all 23
quasars from the BQS sample
(Schmidt \& Green 1983) with $z\le 0.400$, and \nhgal$< 1.9\times10^{20}$ 
cm$^{-2}$, where \nhgal\ is the H~I Galactic 
column density as measured at 21~cm. These selection criteria allow optimal
study of soft X-ray emission at the lowest possible energy. 
The additional advantages of the BQS sample are that it has been extensively
explored at other wavelengths (see Paper I for further details), and
that it includes only bright quasars, thus allowing high S/N X-ray spectra
for most objects. The sample
selection criteria are independent of the quasar's X-ray properties, and
we thus expect our sample to be representative
of the low-redshift, optically-selected quasar population. 

Preliminary results from the analysis of the first 10 quasars available
to us were described by Laor et al. (1994, hereafter Paper I). 
Here we report the analysis of 
the complete sample which allows us to address the three questions posed 
above. The outline of the paper is as follows.
In \S 2 we describe the
observations and the analysis of the spectra. \S 3 describes the
analysis of correlations between the soft X-ray properties and other continuum 
and emission line properties. In \S 4 we compare our results with other 
soft X-ray observations and discuss some of the implications. We conclude 
in \S 5 with answers to the questions raised above, and with some new
questions to be addressed in future studies. 

\section {THE OBSERVATIONS AND ANALYSIS OF THE SPECTRA}

The complete sample of 23 quasars is listed
in Table 1 together with their redshifts, $m_B$ and
$M_B$ magnitudes (calculated for H$_0=50$~km~s$^{-1}$, q=0.5), $R$, 
the radio to optical flux ratio, and 
\nhgal. The redshifts, $m_B$, and $M_B$ 
magnitudes are taken from
Schmidt \& Green (1983), $R$ is taken from Kellermann {\it et al.} (1989).
Note that 4 of the 23 quasars in our sample are radio loud (defined here as 
$R\ge 10$).

The Galaxy becomes optically thick below 0.2~keV for the typical high
Galactic column of $3\times 10^{20}$~cm$^{-2}$ (Dickey \& Lockman 1990; 
Morrison \& McCammon 1983), and 
accurate values of \nhgal\ are therefore crucial even for our low \nhgal\ 
sample. The \nhgal\ values given in column 7 of Table 1 are taken from
Elvis, Lockman \& Wilkes (1989), Savage et al.
(1993), Lockman \& Savage (1995), and the recent extensive
measurements by Murphy et al (1996). All these
measurements of \nhgal\ were made with the 140 foot telescope of the NRAO at
Green Bank, WV, using the ``bootstrapping'' stray
radiation correction method described by Lockman, Jahoda, \& McCammon 
(1986), which provides an angular resolution of 21', and an uncertainty of
$\Delta$\nhgal=$1\times 10^{19}$ cm$^{-2}$ (and possibly lower for our
low \nh\ quasars). This uncertainty introduces a flux
error of 10\% at 0.2~keV, 30\% at 0.15~keV, and nearly a factor of 2 at
0.1~keV. Thus, reasonably accurate fluxes can be obtained down to 
$\sim 0.15$~keV. Note that Murphy et al. (1996) includes accurate \nhgal\
measurements for about 220 AGNs, including most AGNs observed by
{\em ROSAT}, which
would be very useful for eliminating the significant systematic 
uncertainty in the PSPC spectral slope which must be present when
a low accuracy \nhgal\ is used.
 
Table 2 lists the PSPC observations of all the quasars. For the sake of 
completeness
we include also the 10 quasars already reported in Paper I.
All sources were detected, and their net source counts range from 93 to 
 38,015, with a median value
of about 1900 counts. The PROS software package was used to extract
the source counts. Table 2 includes
the exposure times, the dates of the observations, the net number of counts
and their statistical error,
the count rate, 
the radius of the circular aperture used to extract the source 
counts, the offset of the X-ray position from the center of the PSPC field of 
view, the {\it ROSAT} sequence identification number, 
and the SASS version used for the calibration of the data. 
All objects, except one, are 
typically within $15''$ of the center of the PSPC field of view, so all the
identifications are secure. The one exception is PG~1440+356 where the
pointing was offset by $40'$ from the position of the quasar
(Gondhalekar et al. 1994).
Note that the exposure times are uncertain by about 4\% due to a number
of possible systematic errors, as described by Fiore {\it et al.} (1994).
The typically large number of counts for each object allows
an accurate determination of the spectral slope for most objects, as 
described below. 

Model fits to the extracted number of source counts per pulse 
invariant (PI) channel, $N_{\rm ch}^{\rm ob}$, were carried out using the 
XSPEC software package. PI channels 1-12 of the original 256 channels 
spectra ($E<0.11$~keV), were ignored since 
they are not well calibrated and are inherently uncertain due to the
large Galactic optical depth. The January 1993 PSPC calibration matrix was 
used for observations made after 1991 Oct. 14, and earlier observations 
were fit with the March 1992 calibration matrix. The best fit model parameters 
are obtained by $\chi^2$ minimization. Nearby channels were merged when
$N_{\rm ch}^{\rm ob}< 10$. A 1\% error was added in quadrature to the 
statistical
error in $N_{\rm ch}^{\rm ob}$, to take into account possible systematic 
calibration errors (see Paper I for more complete details).

\subsection {A Single Power-Law}

As in Paper I, we fit each spectrum with a single power-law 
of the form $f_E=e^{-N_{\rm H}\sigma_E}f_0E^{\alpha_x}$, where $f_E$
is the flux density,  
$\sigma_E$ is the absorption cross section per H atom (Morrison 
\& McCammon 1983), $f_0$ is the
flux density at 1~keV, and $E$ is in units of keV. 
We make three different fits for each object, with: 
1. \nh\ a free parameter, 2. \nh$=$\nhgal.
3. \nh$=$\nhgal, and $0.47\le E \le 2.5$~keV, i.e. using only the hard 
{\rm ROSAT} band (channels $12-34$ of the rebinned 34 channels 
spectra).
A comparison of fits 1 and 2 allows us to determine 
whether there is evidence for a significant intrinsic absorption excess or
emission excess relative to a single power-law fit with 
\nh$=$\nhgal. 
This comparison also allows us,
as further shown in \S 5.3.3, to determine whether the 21 cm measurement of
\nhgal\ is a reliable measure of the Galactic soft X-ray opacity.  
A comparison of fits 2 and 3 allows us to look for a dependence of 
the power-law slope on energy. 

Table 3 provides the results of fits 1-3 described above. The table
includes the 13 objects not reported in Paper I, and 6 objects from paper
I for which we only now have the accurate \nhgal\ values.
For each fit we give the best fitting spectral slope $\alpha_x$, the 
normalization of the power-law flux at 1 keV ($f_0$), the best fitting
\nh, the $\chi^2$ of the fit ($\chi^2_{\rm fit}$), the number of degrees 
of freedom (dof), and the probability for $\chi^2\ge \chi^2_{\rm fit}$.
The errors $\Delta\alpha_x$ and $\Delta$\nh\ in fit 1 were calculated
by making a grid search for models with $\Delta \chi^2=2.30$, as 
appropriate for 1 $\sigma$ confidence level for two interesting
parameters (e.g. Press {\it et al.} 1989). The error on the slope  
$\Delta\alpha_x$ in fits 2 and 3 is calculated by requiring $\Delta 
\chi^2=1.0$ (i.e. 68\% for one interesting parameter). We neglect the 
effect of $\Delta$\nhgal\ on $\Delta\alpha_x$ in fits 2 and 3 since we use 
accurate \nhgal\ for all objects.

The observed and best fit spectra for the 13 quasars not reported in paper
I are displayed in Figure 1. There are 3 panels for each object. 
The upper panel displays the observed 
count rate per keV as a function of channel energy, the histogram 
represents the expected
rate from the best fit power-law model, $N_{\rm ch}^{\rm ob}$, 
with \nh\ a free parameter (fit 1). 
The middle panel displays $\Delta/\sigma$, where 
$\Delta=N_{\rm ch}^{\rm ob}-N_{\rm ch}^{\rm mod}$, and $\sigma$ is the standard
error in $N_{\rm ch}^{\rm ob}$. This plot helps indicate what features in 
the spectrum 
are significant. The lower panel
displays the fractional deviations from the expected flux, or equivalently 
$\Delta/N_{\rm ch}^{\rm mod}$, which indicates the fractional amplitude of
the observed features. 

As shown in Table 3, in all 13 objects the simple power-law model 
with \nhgal\ (fit 2) provides an acceptable fit (i.e. 
Pr($\chi^2\ge\chi^2_{\rm fit})>0.01$). Note in particular the spectrum of 
PG~1116+215, which despite the very high S/N available (24,272 
net counts), shows no deviations from a simple power law above a level of 
$\lesssim 10$\%. In Paper I a simple power-law model could not provide 
an acceptable fit to three of the 10 quasars,
though in two of them the apparent features could not be fit with a 
simple physical
model, and in one of them this may be due to calibration errors
(see \S 4.1).

As mentioned above, a comparison of the free \nh\ fit (fit 1) with the 
\nhgal\ fit (fit 2) allows us to look for evidence for an absorption 
or an emission excess. We measure the statistical significance of the 
reduction
in $\chi^2_{\rm fit}$ with the addition of \nh\ as a free parameter using
the F test (Bevington 1969). In PG~1440+356 we find a significant
reduction with Pr$=7.5\times 10^{-4}$ 
(F=12.98 for 26 dof), where Pr is the probability that the 
reduction in $\chi^2$ is not statistically significant 
(calculated using the FTEST routine in Press {\it et al.} 1989). The \nh\
obtained in fit 1 suggests intrinsic absorption of about 
$5\times 10^{19}$~cm$^{-2}$ above the Galactic absorption. However,
unlike all other objects, PG~1440+356 was observed significantly off 
axis (see Table 2), and some small systematic calibration 
errors may be present there. Note also that the
$\chi^2_{\rm fit}$ of the fit with \nhgal\ (Pr=0.05) 
is still acceptable. We therefore
cannot conclude that an extra absorber must be present in PG~1440+356.
Marshall et al. (1996) found a very steep slope ($\alpha=-4.7\pm 0.65$)
in PG~1440+35 at 0.1-0.15~keV (80-120\AA) using the Extreme UV Explorer.
There is no indication for such a component in the PSPC spectrum below
0.15~keV (Fig.1c). However,
given the very low sensitivity of the PSPC below 0.15~keV, such a steep
soft component may still be consistent with the PSPC spectrum.
In the other 12 objects the free \nh\ fit does not provide a significant 
improvement (i.e. Pr$>0.01$), and thus there is no clear evidence for either
intrinsic absorption, or low energy excess emission above a simple
power-law. 

Figure 2 compares the Galactic \nh\ deduced from the accurate 21~cm 
measurements with the best fit X-ray column deduced using the free \nh\
fit. The straight line represents equal columns. The $\chi^2$ of the
\nh(21~cm)=\nh(X-ray) model is 31.9 for 22 dof (PG~1001+054 was not
included because of the low S/N), which is acceptable at the 8\% level.
This result demonstrates that there is no significant excess
absorption over the Galactic value in any of our objects. It is interesting 
to note that in our highest S/N spectra, PG~1116+215 and PG~1226+023, 
\nh(X-ray) is determined to a level of $0.8-1\times 10^{19}$~cm$^{2}$, and
is still consistent with \nh(21~cm), indicating that both methods agree to
better than 10\%.

The average hard {\em ROSAT} band (0.5-2~keV) slope for the complete sample is
$-1.59\pm 0.08$ (excluding PG~1114+445 which is affected by a warm absorber,
and PG 1001+054 and PG 1425+267 where the S/N is very low). This slope is
not significantly different from the average slope for
the full {\em ROSAT} band, $-1.63\pm 0.07$. 

Spectral fits to the PSPC data of some of the objects in our complete 
sample have already been reported by Gondhalekar et al. (1994), 
Ulrich \& Molendi (1996), Rachen, Mannheim, \& Biermann 
(1996), and Wang, Brinkmann, \& Bergeron (1996). The results of the 
single power-law fit with a free \nh\ in these papers 
are all consistent with our 
results for the overlapping objects. The only discrepancy is with 
PG~1444+407 where although both us and Wang et al. find a similar slope, 
Wang et al. find evidence for absorption while we find no such 
evidence. For a simple power-law fit to PG~1444+407 Wang et al. 
find $\chi^2=26$ for 20 dof which is acceptable only at the 17\% level 
($\sim 1.4\sigma$ level), while we find for such a fit
$\chi^2=14.7$ for 20 dof, which is acceptable at the 80\% level.

As discussed in paper I (\S 5.1.3), the difference in spectral slopes at hard
(2-10~keV) and soft X-rays raises the possibility that $\alpha_x$ may be 
changing within the PSPC band itself.
 The individual spectra are well fit by a simple power
law, and thus any spectral curvature must be consistent with zero. 
Stronger constraints on the spectral curvature may be obtained by measuring
the average curvature parameter ($\beta$, defined in Paper I) for the
complete sample since the random error in the mean is smaller by 
$\sqrt{N}\sim 5$ than the random error for individual objects. 
 Unfortunately, the PSPC calibration 
uncertainty at low energy, discussed in Paper I, introduces a
systematic error in $\beta$ (which obviously does not cancel out as
$\sqrt{N}$), and as shown in paper I, does not allow a reliable
determination of the curvature parameter.  We therefore did not try to 
constrain the spectral curvature parameter in this paper. 

\section {CORRELATION ANALYSIS}

Table 4 presents 8 of the 12 rest-frame continuum 
parameters and 7 of the 18 emission 
line parameters used for the correlation analysis.
The spectral slopes are defined as
(flux subscript indicates $\log\nu(\rm Hz)$): 
$\alpha_o=\log(f_{15}/f_{14.5})/0.5$
(1-0.3~$\mu$m), $\alpha_{ox}=\log(f_{17.685}/f_{15})/2.685$ 
(3000~\AA-2~keV), and $\alpha_{os}=\log(f_{16.861}/f_{15})/1.861$ 
(3000~\AA-0.3~keV). 
The X-ray continuum parameters are 
from fit 2, and from Paper I. The near IR and optical continuum parameters are 
taken from Neugebauer {\it et al.} (1987). The emission line parameters were
taken from Boroson \& Green (1992). Luminosities were calculated assuming 
$H_0$=50~km~s$^{-1}$ and $q_0=0.5$.

The four additional continuum parameters not presented in Table 5
for the sake of brevity are 
$\alpha_{\rm irx}=\log(f_{16.861}/f_{14.25})/2.611$ (1.69~$\mu$m-2~keV), 
$\alpha_{\rm irs}=\log(f_{17.685}/f_{14.25})/3.435$ (1.69~$\mu$m-0.3~keV),
radio luminosity (Kellermann et al.
1989), and $1~\mu$m luminosity (Neugebauer et al. 1987). The 11
emission line parameters not detailed in Table 5 are: [O~III] EW, 
Fe~II EW, \hb\ EW, He~II EW, [O~III]/\hb\ and He~II/\hb\ flux ratio,
[O~III] peak flux to \hb\ peak flux ratio, radio to optical flux ratio,
and the \hb\ assymetry, shape, and shift parameters. All these 11
parameters are listed in Table 2 of Boroson \& Green (1992). 

The significance of the correlations was tested 
using the Spearman rank-order
correlation coefficient ($r_S$) which is 
sensitive to any monotonic 
relation between the two variables. A summary of
the main correlation coefficients and their two sided significance is given in
Table 5.

\subsection {The Significance Level}

In Paper I the correlation analysis was carried out using 10 objects,
and only relatively strong correlations ($r_S\ge 0.76$) could be detected
at the required significance level (Pr$\le 0.01$). 
Here, with 23 objects, a Pr$\le 0.01$ corresponds to
$r_S\ge 0.52$, and we can thus test for the presence of weaker correlations,
and check whether the correlations
suggested in Paper I remain significant. We have searched for 
correlations among the 12 continuum emission parameters, and between these 12
parameters and the 18 emission line parameters listed above, which gives a 
total of
294 different correlations. One thus expects about 1 spurious correlation
with Pr$\le 3.4\times 10^{-3}$ in our analysis, and for a significance level
of 1\% one would now have to go to Pr$\le 3.4\times 10^{-5}$, rather
than Pr$\le 1\times 10^{-2}$. However, we
find that there are actually 42 correlations with 
Pr$\le 3.4\times 10^{-3}$, rather than just one, in our sample. 
Thus, the probability that any
one of them is the spurious one is only 2.4\%, and the significance
level of these correlations is reduced by a factor of 7 ($=0.024/0.0034$), 
rather than a
factor of 300. Below we assume that correlations with 
Pr$\le 1\times 10^{-3}$ are significant at the 1\% level
(there are 30 correlations with Pr$\le 1\times 10^{-3}$, versus an expected 
number of 0.3). Thus, given the large number of correlations 
we looked at, we can only test reliably for correlations with
$r_S\ge 0.64$ (which corresponds to Pr$\le 1\times 10^{-3}$ for 23 data 
points).

\subsection {The Near IR to X-Ray Energy Distribution}

A comparison of the rest-frame spectral energy distributions of all 23 quasars 
is shown in Figure 3. The 3-0.3~$\mu$m continuum is from Neugebauer et al.
(1987), and the 0.2-2~keV continuum is from paper~I and from this paper.
The upper panel shows the absolute luminosities, and the lower 2 panels the 
luminosity
normalized to unity at $\log \nu=14.25$ for radio quiet quasars and for 
radio loud quasars. Note the relatively small dispersion in the normalized 
0.3~keV ($\log \nu=16.861$) luminosity. 
The outlying objects are labeled. PG~1626+554 is
the only object where a steep $\alpha_x$ is clearly associated with a 
strong soft
excess (relative to the near IR flux). In other objects a steep $\alpha_x$
tends to be associated with a low 2~keV flux. This trend is also suggested 
by the 
presence of a marginally 
significant correlation between $\alpha_x$ and $\alpha_{ox}$ 
($r_S=0.533$, Pr$=0.0089$, see below), and the absence of a significant 
correlation 
between $\alpha_x$ and $\alpha_{os}$ ($r_S=-0.201$, Pr$=0.36$).

The X-ray luminosity distribution appears to be bimodal with two quasars,
PG~1001+054 and PG~1411+ 442, being a
factor of 30 weaker than the mean radio quiet quasar. These two quasars 
appear to form a
distinct group of `X-ray weak quasars'. The statistics for the radio
loud quasars (RLQ) are
much poorer, and there is no well defined mean, but PG~1425+267 may be a 
similar X-ray weak RLQ.

\subsection {Correlations with Emission Line Properties}

Figure 4 presents the correlations between the hard X-ray luminosity, 
$L_{\rm 2\ keV}$ ($\log \nu=17.685$), or the soft X-ray luminosity, 
$L_{\rm 0.3\ keV}$,  with the luminosities of H$_{\beta}$, [O~III], 
He~II, or Fe~II. The value of $r_S$ and the two sided significance
level (Pr) of $r_S$ are indicated above each panel. Upper limits were
not included in the correlations. Thus, the actual correlations for
He~II, where there are 5 upper limits, are likely to be smaller than found
here (there is only one upper limit for [O~III] and Fe~II, and none for
\hb). Excluding He II, the X-ray luminosity is most strongly correlated
with $L_{\rm H\beta}$ ($r_S=0.734$, Pr$=6.6\times 10^{-5}$).
We note in passing that $L_{\rm H\beta}$ has
an even stronger correlation
with the luminosity at 3000~\AA ($r_S=0.866$, Pr$=9\times 10^{-8}$) 
and with the near IR luminosity at 1~$\mu$m ($r_S=0.810$, Pr$=2.7\times 
10^{-6}$).  

The position of the X-ray weak quasars is marked in Fig.4. Both 
PG~1001+054 and PG~1411+442 appear to have an X-ray luminosity weaker
by a factor of about 30 compared to other quasars with similar $L_{\rm H\beta}$.
PG~1425+267 is also weaker by a factor of $\sim 10$ compared with the other
RLQ. These ratios are the same as those found above in \S 3.2, based on the
spectral energy distribution.

Figures 5a-d displays various emission parameters which correlate 
with $\alpha_x$ (as obtained with \nh=\nhgal).
The FWHM of \hb, $L_{\rm [O~III]}$, the Fe~II/H$\beta$ flux ratio,
and the ratio of [O III] peak flux to H$_{\beta}$ peak flux (as
defined by Boroson \& Green) are the emission line parameters which 
correlate most
strongly with $\alpha_x$. As found in Paper I, all the $\alpha_x$ versus
emission line correlations become significantly weaker when we use
$\alpha_x$ obtained with the free \nh\ fit.

The X-ray weak quasars are labeled in the
$\alpha_{\rm ox}$ vs. $\alpha_x$ correlation in Figure 5e. As expected 
they have
a steeper than expected $\alpha_{\rm ox}$ for their $\alpha_x$. The
last parameter shown in Figure 5f is 1.5$L^{1/2}_{14.25}\Delta v^{-2}$, 
where $\Delta v=$\hb\ FWHM. This parameter is related, under some
assumptions, to the luminosity in Eddington units, as further discussed
in \S 4.7.

\section {DISCUSSION}

\subsection {The Soft X-Ray Spectral Shape}

We find an average spectral index $\langle\alpha_x\rangle=-1.62\pm 0.09$ 
for the complete sample of 23 quasars, where 
the error here and below is the uncertainty in the mean. 
This slope is consistent with
the mean slope $\langle\alpha_x\rangle=-1.57\pm 0.06$ which we found
for the subsample of 24 quasars out of the 58 AGNs analyzed by
Walter \& Fink (1993, the other 34 AGNs in their sample are Seyfert
galaxies as defined by V\'{e}ron-Cetty \& V\'{e}ron 1991).
A similar average slope of $-1.65\pm 0.07$ was found by Schartel et al.
(1996) for 72 quasars from the LBQS sample detected in the {\em ROSAT} all
sky survey (RASS). Puchnarewicz et al. (1996) find a significantly
flatter mean slope, $\langle\alpha_x\rangle=-1.07\pm 0.06$, in a large 
sample of 108 soft X-ray selected (0.5-2~keV) AGNs. 
Part of the difference is related to the exclusion of counts below 0.5~keV,
which selects against steep $\alpha_x$ quasars, but this bias cannot explain
the much flatter $\alpha_{ox}$ in their sample. As discussed by 
Puchnarewicz et al., their sample appears to include a large proportion of
highly reddened quasars (see further discussion in \S 4.2).

RLQ are known to have a flatter $\alpha_x$ than radio quiet quasars (RQQ) at  
energies above the PSPC band (e.g. Wilkes \& Elvis 1987, Lawson et al. 1992). 
We find $\langle\alpha_x\rangle=-1.72\pm 0.09$ for the 19 RQQ, and 
$\langle\alpha_x\rangle=-1.15\pm 0.14$ for the 4 RLQ in our sample. We find
a similar trend using the Walter \& Fink quasar data, where 
$\langle\alpha_x\rangle=-1.61\pm 0.08$ for the RQQ
and $\langle\alpha_x\rangle=-1.36\pm 0.08$ for the RLQ. A similar
difference between RQQ and RLQ was found by Ciliegi \& Maccacaro (1996)
in PSPC spectra of a sample of 63 AGNs extracted from the {\em EINSTEIN}
extended medium sensitivity survey sample.
We therefore conclude that the trend observed at harder X-rays also extends 
down to the 0.2-2~keV band. Schartel et al. (1995) stacked PSPC images
of 147 RQQ and 32 RLQ finding for the sum images
$\langle\alpha_x\rangle=-1.65\pm 0.18$ for RQQ
and $\langle\alpha_x\rangle=-1.00\pm 0.28$ for the RLQ. However, the mean
redshift of their objects is $\sim 1.3$ and thus their results apply to
the $\sim 0.45-4.5$~keV band.

As discussed in Paper I, the {\em ROSAT} PSPC indicates a significantly 
different soft X-ray spectral shape for quasars compared with earlier results
obtained by the {\em EINSTEIN } IPC and {\em EXOSAT} LE+ME detectors
(e.g. Wilkes \& Elvis, 1987; Masnou et al. 1992; 
Comastri et al. 1992; Saxton et al. 1993; 
Turner \& Pounds 1989; Kruper, Urry \& Canizares 1990).
In particular, earlier missions suggested that the hard X-ray slope 
(Lawson et al. 1992;  Williams et al. 1992) extends down to 
$\sim 0.5$~keV with a steep rise at lower energy. Here we find that the 
0.2-2~keV spectrum is fit well by a single power-law with Galactic
absorption. This indicates that:
1). the break between the soft and hard X-ray slope must occur well above 
0.5~keV, 2) the break must be gradual, and 3) there is no steep soft
component with significant flux down to $\sim 0.2$~keV. 
ASCA observations of two of the quasars in our sample,
3C~273 by Yaqoob et al. (1994), and PG~1116+215 by Nandra et al. (1996)
find, as expected, significantly flatter spectra above 2~keV. However,
the exact break energy cannot be accurately determined from the ASCA
spectra due to likely calibration uncertainties below 1~keV.

As mentioned in Paper I, the different {\em EINSTEIN } IPC and 
{\em EXOSAT} LE+ME results may be 
traced back to the combined effect of the lower sensitivity of these
instruments below $\sim 0.5$~keV, and possibly
some calibration errors. Small systematic errors in the PSPC response function 
appear to be present below 0.2~keV (Fiore {\it et al.} 1994), and this
instrument is thought to be significantly better calibrated at
low energy than earlier instruments.

No significant spectral features are present in the PSPC spectra of all 13 
additional quasars reported here, indicating that intrinsic features must 
have an
amplitude of less than 10-20\%. Note, in particular the high S/N spectrum 
of PG~1116+215, where the number of counts is about 12 times the median
sample counts, yet this spectrum is still consistent with a simple
power-law. For the complete sample we find that only 1 quasar, PG 1114+445,
has a significant physical feature which is well described by a warm 
absorber model. In two other quasars,
PG~1226+023 and PG~1512+370, there are significant features 
below 0.5~keV (paper I). In the case of PG~1512+370 the features are at a 
level of $\sim 30$\%, and in PG~1226+023 they are at a level $\lesssim$10\% 
and may well
be due to small calibration errors. This result is consistent with the
result of Fiore et al. (1994) who found that a simple power-law
provides an acceptable fit to the individual spectra of six high S/N 
PSPC quasar spectra. 

A composite optical to hard X-ray spectral energy distribution for RLQ and
RQQ is displayed in Figure 6. To construct it we used the mean $L_{14.25}$
(Table 5), the mean $\alpha_o$, the mean $\alpha_x$, 
and the mean $\alpha_{ox}$ in 
our sample. We excluded from the mean the three X-ray weak quasars, and
PG~1114+445, where $\alpha_x$ is highly uncertain due to the presence
of a warm absorber. The mean spectra were extended above 2~keV assuming a slope of
$-1$ for RQQ and $-0.7$ for RLQ.
The Mathews \& Ferland (1987, hereafter MF) quasar 
energy distribution is also displayed for the purpose of comparison. The
MF shape assumes a steep soft component with a break to the hard X-ray slope
above 0.3~keV, and it therefore significantly underestimates the soft X-ray
flux at $\sim 0.2-1$~keV.

RLQ tend to be somewhat stronger hard X-ray sources than RQQ. 
This trend, together with the flatter X-ray slope of RLQ was interpreted by 
Wilkes \& Elvis (1987) as possible evidence for a two component model. In this 
interpretation RLQ have the same hard X-ray component with
$\alpha_x\sim -1$, as in RQQ, with an additional contribution from a
flatter 
$\alpha_x\sim -0.5$ component, making their overall X-ray emission
flatter and brighter. The additional X-ray component in RLQ could be related
to the radio jet, e.g. through inverse Compton scattering. The composite
spectrum suggests that although RLQ
are brighter at 2~keV, they may actually be fainter at lower energy because
of their flatter $\alpha_x$. The RLQ composite is based only on four 
objects and is therefore rather uncertain. In addition, the results of Sanders
et al. (1989 \S III.c) suggest that RLQ in the PG sample 
are about twice as bright at 2~keV
compared with RQQ of similar optical luminosity, rather than 
the $\sim 30$\% found for the composite, thus the difference in PSPC $\alpha_x$
would imply a smaller difference at 0.2~keV than shown in the composite.
If RLQ are indeed weaker than RQQ at 0.2~keV then the two component model 
suggested above would not be valid, and RLQ need to have a different X-ray 
emission process, rather than just an additional component. 

The difference between RLQ and RQQ may actually be unrelated to the
radio emission properties, as discussed in \S 4.4.

Figure 6 also displays a simple cutoff power-law model of the form 
$ L_{\nu}\propto \nu^{\alpha_o}e^{-h\nu/kT_{\rm cut}}$ with $\alpha_o=-0.3$ 
and $T_{\rm cut}=5.4\times 10^5$~K. This is an 
alternative way to interpolate between the UV and soft X-ray emission, 
and it is also
a reasonable approximation for an optically thick thermal component. 
The lack of a very steep low energy component down to 0.2~keV 
allows us to
set an upper limit on $T_{\rm cut}$. The upper limit is set using 
$\alpha_o$ and $\alpha_{os'}$ the slope from 3000~\AA\ to rest frame 
$0.15(1+z)$~keV
(the lowest energy where the Galactic absorption correction error$\le30$\%),
given by 
\[ \alpha_{os'}=[2.685\alpha_{ox}-(17.685-\log\nu_{s'})\alpha_x]/
\log (\nu_{s'}/10^{15}), \] 
\[ \ \ {\rm where}\ \  
\log \nu_{s'}=16.560+\log(1+z). \]
The upper limit on the cutoff temperature $T_{\rm cut}^{\rm ul}$ is related to
the spectral slopes by 
\[ 
T_{\rm cut}^{\rm ul}=\frac{
4.8\times 10^{-11}(\nu_{s'}-10^{15})\log e}{
(\alpha_{o}-\alpha_{os'})\log(\nu_{s'}/10^{15})}~{\rm K}. 
\]
We find a rather small dispersion in $T_{\rm cut}^{\rm ul}$   
with $\langle T_{\rm cut}^{\rm ul} \rangle=(5.5\pm 2.6)\times 10^5$~K, 
averaged over
the complete sample (Table 4), which corresponds to a cutoff energy of 
47~eV, or about
3.5 Ryd. This value of $T_{\rm cut}^{\rm ul}$ corresponds very closely to the
far UV continuum shape assumed by MF (see Fig.6).

Walter et al. (1994) fit such a cutoff model directly to
six quasars and Seyfert galaxies finding
$\langle E_{\rm cut}\rangle=63\pm 12$~eV, or  
$\langle T_{\rm cut}\rangle=(7.3\pm 1.4)\times 10^5$~K, while
Rachen, Mannheim \& Biermann (1996) find using such a model
$\langle T_{\rm cut}\rangle=(6.3\pm 2.3)\times 10^5$~K for 7 
quasars and Seyfert galaxies. These values are
consistent with our results. The small dispersion in $T_{\rm cut}$
reflects the small dispersion in $\alpha_{\rm os}$ in our sample, which
is in marked contradiction with the dispersion predicted by thin accretion
disk models, as further discussed in \S 4.5.

\subsubsection{The Far UV Continuum}

Zheng et al. (1996) have constructed a composite quasar spectrum 
based on HST spectra of 101 quasars at $z>0.33$. They find a
far UV (FUV) slope (1050\AA-350\AA) of 
$\langle \alpha_{\rm FUV}\rangle=-1.77\pm 0.03$ for RQQ and 
$\langle \alpha_{\rm FUV}\rangle=-2.16\pm 0.03$ for RLQ, with slopes
of $\sim -1$ in the 2000\AA-1050\AA\ regime. The Zheng et al. mean 
spectra, presented in Fig.6, together with the PSPC mean spectra, 
suggest that the FUV power-law
continuum extends to the soft X-ray band. In the case of RQQ there
is remarkable agreement in both slope and normalization of the soft
X-ray and the FUV power-law
continua. RLQ are predicted to be weaker than RQQ at $\sim 100$~eV
by both the FUV and the PSPC composites. It thus appears that there
is no extreme UV sharp cutoff in quasars, and that the fraction
of bolometric luminosity in the FUV regime is significantly smaller
than assumed. 

The UV to X-ray continuum suggested in Fig.6 is very different from the
one predicted by thin accretion disk models (\S 4.5), and suggested
by photoionization models. In particular, it implies about four times
weaker FUV ionizing continuum compared with the MF continuum which
was deduced based on the
He~II~$\lambda 1640$ recombination line.

One should note, however, that
the Zheng et al. sample is practically disjoint from our low $z$ sample,
so it may still be possible that low $z$ quasars have a different FUV
continuum.

\subsection {Intrinsic Absorption}

As shown in Fig.2, the H~I column deduced from our accurate 21~cm
measurements is consistent for all objects with the best fit X-ray column. 
It is quite remarkable that even in our highest S/N spectra the 21~cm
and X-ray columns agree to a level of about $1\times 10^{19}$~cm$^{-2}$, 
or 5-7\%. 
This agreement is remarkable since the 21~cm line and the PSPC are actually
measuring the columns of different elements. Most of the soft X-ray 
absorption is due to He~I or He~II, rather than H~I, and the H~I column
is indirectly inferred assuming the column ratio H~I/He~I$=10$. 
The fact that the 21~cm line and the PSPC give the same H~I 
column implies that the H~I/He column ratio at high Galactic latitudes
must indeed be close to 10. The dispersion in the H~I/He column ratio
is lower than 20\% (based on typical quasars in
our sample), and may even be lower than 5\% (based on our highest S/N 
spectra). There is therefore no appreciable Galactic column at high
Galactic latitudes where the
ionized fraction of H differs significantly from the ionized fraction
of He, as found for example in H~II regions (e.g. Osterbrock 1989).

The consistency of the 21~cm and X-ray columns also indicates 
that the typical column of cold gas intrinsic to the quasars in our sample 
must be
smaller than the X-ray \nh\ uncertainty, or about 
$3\times 10^{19}(1+z)^3$~cm$^{-2}$.
An additional indication for a lack of an intrinsic cold column in quasars
comes from the fact the the strong correlations of $\alpha_x$ 
with the emission line parameters described above (\S 3.3) become weaker
when we use $\alpha_x$ from the free \nh\ fit rather than $\alpha_x$ from
the fit with \nhgal. This indicates that \nhgal\
is closer to the true \nh\ than the free \nh\ (see discussion in Paper I).
In our highest S/N spectra we can set an upper limit of 
$\sim2\times 10^{19}$~cm$^{-2}$ on any intrinsic absorption. 
As discussed in Paper I, the lack of intrinsic X-ray column for most quasars 
is consistent with more stringent
upper limits set by the lack of a Lyman limit edge, as well as the He~I and
the He~II bound-free edges in a few very high z quasars.

Puchnarewicz et al. (1996) suggest that the strong $\alpha_x$ vs. 
$\alpha_{ox}$ correlation in their X-ray selected sample is due to absorption 
of the optical
and soft X-ray emission by cold gas and dust. They show that the $\alpha_x$ 
vs. $\alpha_{ox}$ correlation for the 10 objects in Paper I can be explained
by a universal spectral shape absorbed by a gas with a column of up to
\nh$=3\times 10^{20}$~cm$^{-2}$ (see their Figure 16). As described above, 
such absorbing columns are clearly ruled out by our high S/N spectra.

The mean $\alpha_o$ in the Puchnarewicz et al. sample is $-0.92\pm 0.07$,
which is significantly steeper than the mean $\alpha_o$ for optically
selected quasars, e.g. a median of $-0.2$ for 105 PG quasars 
(Neugebauer et al. 1987), a median of $-0.32$ for 718 LBQS quasars
(Francis et al. 1991), and $\langle\alpha_o\rangle=-0.36\pm 0.05$, 
in our sample. Puchnarewicz et al. suggested that the much flatter 
$\alpha_o$ of the PG quasars is a selection bias since these quasars were
selected by the strength of their UV excess. However, the PG sample was
selected on the basis of the color criterion $U-B<-0.44$, which using
the flux transformations of Allen (1973), corresponds to $\alpha_o\ge-1.8$.
Thus, most of the red quasars discovered by Puchnarewicz et al. fit into
the PG color criterion. The difference between the soft X-ray selected and
optically selected quasars must reflect the true tendency of 
quasars selected above 0.5~keV 
to be significantly redder than optically selected quasars. 
These red quasars may very well be affected by a large 
absorbing column (\nh$>10^{21}$~cm$^{-2}$), as suggested by 
Puchnarewicz et al. 

Intrinsic absorption is common in Seyfert 1 galaxies. About half of the
primarily X-ray selected
Seyfert galaxies observed by Turner \& Pounds (1989) using the {\em EXOSAT}
LE+ME detectors, 
by Turner, George \& Mushotzky (1993) using the {\em ROSAT} PSPC, and by 
Nandra \& Pounds (1994) using the {\em GINGA} LAC for a largely overlapping 
sample, show low energy 
absorption, or spectral features inconsistent with the simple power-law 
typically observed above 2~keV. Quasars are very different. Excess
absorption produces significant spectral features only in one object
(PG~1114+445, see paper~I), i.e. $\sim 5$\% 
(1/23) of the objects, and the absorbing gas is partially ionized ("warm"), 
rather than neutral. Given the typical S/N in our sample we estimate that
a partially ionized absorber which produces $\tau>0.3$ can be ruled out
in most of our objects.
We cannot rule out partial
absorption, or complete absorption and scattering,
by a very high column density (\nh$>10^{24}$~cm$^{-2}$) gas since such
effects may only suppress the flux level without affecting
the spectral shape, and without inducing significant spectral features. 
As described in \S 4.8, we suspect that such strong absorption may indeed be 
present in about $\sim 10$\% (3/23) of the quasars in our sample (the X-ray 
weak quasars).

\subsection {Implications of the Continuum-Continuum Correlations}

The continuum-continuum luminosity correlations found here are
all weaker than found in Paper I. This is mostly due to the three X-ray weak
quasars which were not present in Paper I. For example, in Paper I we 
found that $f_{0.3~{\rm keV}}$ can be predicted to within a
factor of two, once $f_{1.69~\mu m}$ is given. This statement is still valid
if the 4 extreme objects labeled in Fig.3 middle panel are excluded.
The implications of the near IR versus X-ray luminosity correlation on the
X-ray variability power spectrum were discussed in Paper I.  

In Paper I we noted the similarity
$\langle\alpha_{ox}\rangle=\langle\alpha_x\rangle=-1.50$, which was also
noted by Brunner {\it et al.} (1992) and Turner, George \& Mushotzky (1993).
However, we argued there that this similarity is only fortuitous, and 
that it does not imply that the X-ray power law can be extrapolated into
the UV since the optical slope is significantly different. Here we find that 
the relation 
$\langle\alpha_{ox}\rangle\simeq \langle\alpha_x\rangle$ holds only
roughly for the complete sample where 
$\langle\alpha_{ox}\rangle=-1.55\pm 0.24$, and
$\langle\alpha_x\rangle=-1.62\pm 0.45$. This relation 
does not hold when the sample is broken to the RQQ where 
$\langle\alpha_{ox}\rangle=-1.56\pm 0.26$, and 
$\langle\alpha_x\rangle=-1.72\pm 0.41$, 
and to the RLQ where 
$\langle\alpha_{ox}\rangle=-1.51\pm 0.16$, and
$\langle\alpha_x\rangle=-1.15\pm 0.27$.

A significantly flatter $\langle\alpha_{ox}\rangle$ is obtained when the
three X-ray weak quasars, and the absorbed quasar PG~1114+445 are excluded.
Thus, ``normal'' RQQ quasars in our sample have
$\langle\alpha_{ox}\rangle=-1.48\pm 0.10$, $\langle\alpha_x\rangle=-1.69\pm 
0.27$, while for the RLQ
$\langle\alpha_{ox}\rangle=-1.44\pm 0.12$, $\langle\alpha_x\rangle=-1.22\pm 
0.28$, where the $\pm$ denotes here and above the dispersion about the mean,
rather than the error in the mean.

The $\alpha_x$ versus $\alpha_{ox}$ correlation found here is weaker than in
Paper I due to the presence of the X-ray weak quasars. However, the other
20 quasars appear to follow a trend of increasing $\alpha_x$ with
increasing $\alpha_{ox}$ (Fig.5e), indicating as discussed in Paper I
that a steep $\alpha_x$ is generally associated with a weak hard X-ray
component (at 2~keV), rather than a strong soft excess. The only object
which clearly violates this trend is PG~1626+554 (Fig.3), which has both
a steep $\alpha_x$ and a strong soft excess. Puchnarewicz, Mason \& Cordova
(1994) and Puchnarewicz et al. (1995a; 
1995b) present PSPC spectra of three AGNs with extremely strong soft excess,
where $L_{\rm 0.2~keV}>L_{\rm 3000~\AA}$. Our sample suggests that such objects
are most likely rare, as can also be inferred from the selection criteria of
Puchnarewicz et al. who selected their three objects from  
the {\em ROSAT} WFC all sky survey, in which only five AGNs were detected.
This selection criterion implies that these AGNs must 
have a very high far UV flux.

The soft X-ray selected quasars in the Puchnarewicz et al. (1996) sample
have $\langle\alpha_{ox}\rangle=-1.14\pm 0.02$, and none of their quasars
is ``X-ray weak'', i.e with $\alpha_{ox}<-1.6$. The absence of X-ray
weak quasars in their sample is clearly a selection effect. The small
survey area implies that most quasars in their sample are optically rather
faint ($m_B\sim 18$-19). ``Normal'' $\alpha_{ox}$ quasars in their sample 
produce a few hundred PSPC counts, but ``X-ray weak'' quasars are below 
their detection threshold. The abundance of ``X-ray loud'' quasars 
(i.e $\alpha_{ox}>-1$) in the Puchnarewicz et al. sample is consistent 
with their rarity in optically selected samples. For example, 
quasars with $\alpha_{ox}=-1$ are about 20 times fainter at 3000\AA\
than quasars with $\alpha_{ox}=-1.5$, for the same $L_x$. Since the
surface density of quasars increases as $\sim f_{\rm B}^{-2.2}$ (see
\S 2.2.2.1 and Figure 1 in Hartwick \& Schade), where 
$f_{\rm B}$ is the B band flux,
there are about 700 times more of these fainter quasars per B magnitude
per square degree. Thus, even if only 0.3\% of quasars at a given 
$f_{\rm B}$
have  $\alpha_{ox}=-1$, there would still be twice as many quasars with
$\alpha_{ox}=-1$
than $\alpha_{ox}=-1.5$ per square degree in an X-ray flux limited sample,
such as the Puchnarewicz et al. sample.

\subsection {Implications of The Continuum-Line Correlations} 

The presence of the strong correlations of $\alpha_x$ with the \hb\ FWHM,
with $L_{\rm [O~III]}$, and with the Fe~II/\hb\ ratio described in
Paper I is verified here. The correlation coefficients for the complete
sample are comparable or somewhat smaller than those found in Paper I, but
since the sample is larger the significance level is now much higher
(Fig.5). We also report here an additional strong correlation of
$\alpha_x$ with the ratio of [O~III] peak flux to the \hb\ peak flux,
which is one of the emission line parameters measured by Boroson \& Green.
The $\alpha_x$ versus \hb\ FWHM correlations is
the strongest correlation we find between any of the X-ray continuum 
emission parameters and any of the emission line parameters reported by
Boroson \& Green (with the addition of line luminosities reported in Table 5).

The $\alpha_x$ versus \hb\ FWHM correlation is much stronger than
the well known $\alpha_x$ correlation with radio loudness  
($r_S=0.26$, Pr=0.23 in our sample, but see Wilkes \& Elvis 1987 and 
Shastri et al. 1993 for stronger correlations). Thus, the fact that
the average $\alpha_x$ in RLQ is significantly flatter than in
RQQ (\S 4.1) may be completely unrelated to the presence of radio 
emission, it may just reflect
the fact that RLQ tend to have broader lines than RQQ (e.g. Tables 3 and
5 in Boroson \& Green). This
appears to be the case in our sample, where the RLQ follow the same
$\alpha_x$ versus \hb\ FWHM distribution defined by the RQQ (Fig.5a). 
This intriguing suggestion can be clearly
tested by comparing $\alpha_x$ for RLQ and RQQ of similar \hb\ FWHM.

Boller, Brandt \& Fink (1996) studied in detail narrow line Seyfert 1 
galaxies (NLS1) and they also find an apparently significant trend of
increasing $\alpha_x$ with increasing  \hb\ FWHM. However, the scatter in
their sample is significantly larger than that found here. In particular
they find a large range of $\alpha_x$ for \hb\ FWHM$<2000$~km~s$^{-1}$,
where only a few objects are available in our sample.
The overall larger scatter in the Boller, Brandt \& Fink data is probably 
due in part to the generally larger statistical 
errors in their $\alpha_x$ determinations. Large
systematic errors may also be induced by the use of \hb\ FWHM from
a variety of sources. The measured \hb\ FWHM can be sensitive to the
exact measuring procedure, such as continuum placement, subtraction
of Fe~II blends, and subtraction of the narrow component of the line 
(produced in the narrow line region) which may not be well resolved
in low resolution spectra. For example, Shastri et al. 1993 and 
Boroson \& Green 
measured the \hb\ FWHM independently for 13 overlapping objects, in 8
of which their values differ by more than 1000~km~s$^{-1}$.
Other than these technical reasons the
increased scatter may represent a real drop in the strength of the 
correlation when the luminosity decreases from the quasar level studied here
to the Seyfert level studied by Boller, Brandt \& Fink. One should also
note that intrinsic absorption is probably common in Seyfert 1 galaxies 
(Turner \& Pounds, 1989; Turner, George \& Mushotzky, 1993), and such an 
absorption may lead to a large 
systematic error in $\alpha_x$ unless a high S/N spectrum is available
indicating that features are present. 

Wang, Brinkmann \& Bergeron (1996) analyzed PSPC spectra of 86 AGNs, 
including 22 of the 23 quasars from our sample. Their sample is more
heterogeneous than ours and includes some high z quasars and a number
of AGNs selected by their strong Fe~II emission. The various correlations
found by Wang et al. are typically similar, or somewhat weaker than found
here. For example, their(our) values are $r_S=-0.73(-0.79)$ for 
$\alpha_x$ versus \hb\ FWHM, and $r_S=0.65(0.714)$ for $\alpha_x$ versus 
Fe~II/\hb. The somewhat smaller values found by Wang et al. may result
from their inclusion of $z>0.4$ quasars, where $\alpha_x$ measures a 
higher energy slope than measured here, and from their use of a free \nh\
fit (limited from below by \nhgal), which increases the random error
in $\alpha_x$ (see Paper I).

We verify the strong correlation between $L_{\rm H\beta}$ and 
$L_{\rm 2\ keV}$ found in Paper I. The correlations of the other lines 
with X-ray luminosity are significantly weaker than found in Paper I, 
and they are only marginally significant.
Corbin (1993) found significant correlations of $L_{\rm 2\ keV}$ with 
Fe~II/\hb\ ($r_S=-0.474$), and of $L_{\rm 2\ keV}$ with the \hb\ asymmetry 
($r_S=-0.471$). We find that neither correlation is significant in our 
sample ($r_S=-0.288$, Pr=0.19, and $r_S=-0.106$, Pr=0.63). Since
we can only test for correlation with $r_S>0.64$, we cannot securely 
exclude the presence of the correlations reported by Corbin.

As discussed in Paper I, the $\alpha_x$ versus $L_{\rm [O~III]}$ 
correlation
can be used to place a limit on the $\alpha_x$ variability on timescales
shorter than a few years. Given the scatter in this correlation we estimate
that $\alpha_x$ should not vary by significantly more than 0.3 on these
timescales.

It is hard to interpret the $\alpha_x$ versus [O~III] to \hb\ peak flux 
ratio correlation since the physical meaning of the [O~III] to \hb\ peak 
flux ratio parameter defined by Boroson \& Green is rather obscure. 
The [O~III] peak flux is related to the width of [O~III], and the
[O~III] to \hb\ peak flux ratio may thus partly reflect the FWHM ratio of 
these lines. Thus, this correlation may represent a correlation of the
[O~III] FWHM with $\alpha_x$. High spectral resolution measurements of the
[O~III] line profile are required to test this possibility. 

\subsection {Inconsistency with Thin Accretion Disk Models} 

Figure 7 presents the continuum emission from two thin accretion disk 
models. The models are for a disk around a rotating
black hole, and viscous stress which scales like the 
$\sqrt{P_{\rm gas}P_{\rm rad}}$, where $P_{\rm rad}$ is the radiation 
pressure and $P_{\rm gas}$ is the gas pressure
(Laor \& Netzer 1989).
Significant soft X-ray emission is obtained for disks
with a high accretion rate and a small inclination. However, as discussed
by Fiore et al. (1995), the observed
soft X-ray spectral slope is always much flatter than the one produced 
by a thin `bare' accretion disk model. As noted above there is no 
indication in the 0.2-2~keV band for a very steep and soft ``accretion disk''
component. 

Although thin disks cannot
reproduce the 0.2-2~keV spectral shape, they may still be able
to contribute a significant fraction of the flux at the lowest
observed energy, i.e. 0.2-0.3~keV, above which a non thermal power-law
component sets in. As noted by Walter et al. (1994) and in Paper I, 
accretion disk models
predict a large dispersion in the optical/soft X-ray flux ratio, and
the strong correlation between these fluxes argues against 
the idea that a thin disk produces both the optical and
soft X-ray emission. The arguments put by Walter et al. and
in Paper I were only qualitative, and were not based on actual disk
models. Furthermore, the objects in the small sample of Walter et al.
were selected from known optically bright AGNs, and they also had to
be bright soft X-ray sources since most spectra were obtained from the 
{\em ROSAT} all sky survey. Thus, these objects were a priori selected
to be bright at both optical-UV and at soft X-rays, and the absence of
a large scatter in the UV/soft X-ray flux ratio may just reflect the sample
selection criteria. Such selection effects are not present in our sample
since the sample was defined independently of the X-ray properties, and 
X-ray spectra were obtained for all objects. 

Below we describe a detailed 
calculation of the expected distribution of optical/soft X-ray flux
ratio for a complete optically selected sample 
based on the thin disk models of Laor \& Netzer (1989), and
show that such models cannot be reconciled with the observed 
distribution of optical/soft X-ray flux ratio in our complete sample.

The optical/soft X-ray flux ratio, $\alpha_{os}$, of a given disk
model depends on the black hole mass, accretion rate $\dot{m}$, and
inclination angle $\theta=\cos^{-1}\mu$. We now need to determine what
distributions of these parameters will be consistent with the
observed luminosity function in a complete optically selected sample.  
The intrinsic distribution of disk inclinations must be random. However,
the observed distribution depends on the shape of the luminosity 
function of quasars, and possible obscuration effects, as described below.

The luminosity function of quasars is parametrized using the number
density of quasars per unit volume per magnitude 
$\Phi\equiv d^2N/dMdV$, and it is well fit by a power-law  
over a restricted range of magnitude, M. Using Figure 2 in Hartwick \&
Schade (1990) we find $\log \Phi= 0.55M+c$ for $z<0.2$ and
$\log \Phi= 0.66M+c$ for $0.4<z<0.7$, where $c$ is a constant. 
Since our sample is 
restricted to $z<0.4$ we assume $\log \Phi= 0.6M+c$. Using the
relation $M=-2.5\log L+c$ we get that $dn/dL\propto L^{-2.5}$, where
$n\equiv dN/dV$. 

The apparent luminosity of a flat disk $L_{\rm app}$ is related to its
intrinsic luminosity through $L_{\rm app}=2\mu L$, neglecting limb
darkening effects which steepen the $\mu$ dependence, and relativistic
effects which flatten the $\mu$ dependence. This provides a reasonable
approximation in the optical-near UV regime (see Laor, Netzer, \& Piran 
1990).
Assuming $dN/d\mu=const$, i.e. a uniform distribution of inclination
angles for the intrinsic quasar population, we would like to find 
$dn/d\mu$ for a given $L_{\rm app}$.
When $L_{\rm app}$ is fixed, $\mu \propto L^{-1}$, and substituting $\mu$
in the expression for $dn/dL$ we get $dn/d\mu\propto \mu^{0.5}$. Thus
although the disks are assumed to have a uniform distribution of 
inclination angles the observed distribution at a  given $L_{\rm app}$ 
is biased towards face on disks.

To reproduce the observed luminosity function we choose two
values of $\dot{m}=0.1, 0.3$, where $\dot{m}$ is measured in units of the
Eddington accretion rate. Since $L\propto m_9\dot{m}$, where
$m_9$ is the black hole mass in units of $10^9M_\odot$
the required mass distribution is $dn/dm_9\propto m_9^{-2.5}$. 
The observed number of objects in a flux limited sample
is $dN_{\rm ob}/dL\propto dn/dL\times V(L)$, where 
$V(L)\propto L^{3/2}$ is the observable volume for a flux limited 
sample, such as the BQS sample. We therefore select a mass distribution
of $dN_{\rm ob}/dm_9\propto m_9^{-1}$.

Figure 8 compares the observed distribution of $\alpha_{os}$, as a 
function of $\nu L_{\nu}$ at 3000\AA, with the one expected from thin 
accretion disk 
models with the parameter distribution described above. 
Thin disk models cannot account for the very small scatter in 
$\alpha_{os}$.

The range of observed disk inclinations may actually be smaller than
assumed here. For example, for a certain range of inclinations the
disk may be completely obscured by an optically thick torus, as
suggested in unification schemes for RQQ (e.g. Antonucci 1993). However,
even if $\mu$ is fixed at a given value for all AGNs (say $\mu=1$ which
corresponds to the points extending from $\alpha_{os}=-1.5$ on the
left axis to $\log \nu L_{\nu}=46.5$ on the bottom axis of Fig.8),
the range in $m_9$ and $\dot{m}$  will stil produce a range in $\alpha_{os}$
which is much larger than observed. 

The X-ray power-law emission is most likely produced by Comptonization
of the thermal disk emission in a hot corona above the disk (e.g. Czerny
\& Elvis 1987). The slope and normalization of the power-law component
are determined by the temperature and electron scattering optical depth
in the corona (e.g. Sunyaev \& Titarchuk 1985; Titarchuk \& Lyubarskij 1995).  
The small range in $\alpha_{os}$
implies that some physical mechanism which couples the optical and
soft X-ray emission processes must be operating, e.g. through a
feedback which regulates both the temperature (see Haardt \& Maraschi 1993)
and the optical depth of the corona. 

As pointed out by various authors (Ross, Fabian \& Mineshige, 1992; 
Shimura \& Takahara 1995;
Dorrer et al. 1996), and shown in Fig.7, simple thin accretion disks with no
corona can produce a significant flux below 1~keV. For various disk model 
parameters $\alpha_{os}$ can in fact be significantly flatter than observed 
(Fig.8), yet such extreme flat optical-soft X-ray spectra are only rarely
observed (e.g. Puchnarewicz 1995a). The flattest spectra are expected for disks
which are close to edge on (e.g. Laor, Netzer \& Piran 1990), and one therefore
needs to assume that such disks are not observable. This is indeed expected in
AGNs unification schemes which invoke obscuring material close to the disk
plane. Alternatively, the accreted material may form a geometrically thick,
rather than a thin, 
configuration close to the center, which would display a smaller inclination
dependence. 

\subsection {Inconsistency with Optically Thin Free-Free 
Emission Models}

As was clearly demonstrated by Fiore et al. (1995) for 6 low redshift
quasars with a high S/N PSPC spectra, and by
Siemiginowska et al. (1995) using {\em EINSTEIN} data for 47 quasars
from Elvis et al. (1994), isothermal optically thin pure free-free emission
models (Barvainis 1993) cannot fit the
observed UV to soft X-ray energy distribution in AGNs. Furthermore, as was
pointed out by Kriss (1995), and Hamman et al. (1995), optically thin free-free
emission can also be ruled out based on the observed UV line emission.

\subsection {On the origin of the $\alpha_x$ versus \hb\ FWHM
correlation}

What is the physical process behind the $\alpha_x$ versus \hb\ FWHM
correlation?  In Paper I we speculated that this may either be an
inclination effect, or that it could be an $L/L_{\rm Edd}$ effect.
Various authors raised the interesting suggestion that steep
$\alpha_x$ quasars may be analogous to `high'-state Galactic black
hole candidates (e.g. White, Fabian \& Mushotzky 1984; 
Fiore \& Elvis 1995; Pounds, Done \& Osborne 1995), which display a 
steep slope in the soft and the hard X-ray bands when 
their brightness increases. The physical interpretation for this effect
described by Pounds, Done \& Osborne (1995) is that the hard X-ray 
power-law is produced by Comptonization in a
hot corona and that as the object becomes brighter in the optical-UV,
Compton cooling of the corona increases, the corona becomes colder,
thus producing a steeper X-ray power-law. This is obviously far from
being a predictive model since the corona heating mechanism
is not specified, and it is implicitly assumed that the coronal heating 
does not increase much as the quasar becomes brighter. However, the narrow
\hb\ line profiles provide independent evidence that steep $\alpha_x$
quasars may indeed have a higher $L/L_{\rm Edd}$, as further
described below.

The $L/L_{\rm Edd}$ of quasars can be estimated under two
assumption: 1. The bulk motion of the gas in the broad line region is 
virialized, i.e. $\Delta v\simeq \sqrt{GM/r}$, 
where $\Delta v$=\hb\ FWHM. This gives 
\[ \Delta v_{3000}=2.19m_9^{1/2}R_{0.1}^{-1/2},\]
 where
$\Delta v_{3000}=\Delta v/3000$~km~s$^{-1}$, $m_9=M/10^9~M_{\odot}$, and
$R_{0.1}=R/0.1$~pc. 2. The size of the broad line region is
determined uniquely by the luminosity, $R_{0.1}=L_{46}^{1/2}$~pc,
where $L_{46}=L_{\rm Bolometric}/10^{46}$. This
scaling is consistent with reverberation line mapping of AGNs
(Peterson 1993; Maoz 1995), and is theoretically expected if the
gas in quasars is dusty (Laor \& Draine 1993, Netzer \& Laor 1994).
Combining assumptions 1 and 2 gives 
\[ \Delta v_{3000}=2.19m_9^{1/2}L_{46}^{-1/4}, \]
and thus the mass
of the central black hole is 
\[ m_9=0.21\Delta v_{3000}^2L_{46}^{1/2}. \]
Using $L_{\rm Edd,46}=12.5m_9$ one gets
\[ L/L_{\rm Edd}=0.38\Delta v_{3000}^{-2}L_{46}^{1/2}. \]
Thus, given the two assumptions made above, narrow line quasars should 
indeed have a high $L/L_{\rm Edd}$, 
as previously suggested based only on their steep 
$\alpha_x$, and analogy to Galactic black hole candidates.

To test whether $L/L_{\rm Edd}$ is indeed the underlying parameter
which determines $\alpha_x$, rather than just the \hb\ FWHM, we
looked at the correlation of $\alpha_x$ versus 
$\Delta v_{3000}^{-2}L_{46}^{1/2}$ displayed in Fig.5f, where we used the
1.7~$\mu$m luminosity and the relation $L_{\rm Bolometric}=15L_{14.25}$
(see Fig.7 in Laor \& Draine). This correlation is not as strong as the
$\alpha_x$ versus \hb\ FWHM correlation, but it certainly appears 
suggestive. Note that $L/L_{\rm Edd}>1$ for some of the objects in
the sample. These values are well above the thin accretion disk limit 
($L/L_{\rm Edd}=0.3$, Laor \& Netzer 1989) and suggest a thick disk 
configuration. However, the
assumptions used above to infer $L/L_{\rm Edd}$ are more qualitative
than quantitative since both the luminosity and the velocity field in the
broad line region may not
be isotropic and therefore the presence of $L/L_{\rm Edd}>1$ 
cannot be securely deduced.  

The Pounds et al. mechanism implies that a steep $\alpha_x$ is
associated with a weak hard X-ray component, and as described in
\S 3.2, this indeed appears to be the trend in our 
sample. If the Pounds et al. mechanism is true then steep $\alpha_x$
AGNs should also have a steep hard X-ray power-law. We are currently
pursuing this line of research using ASCA and SAX observations of
our sample.

An additional hint that a steep $\alpha_x$ may indeed be associated
with a high 
$L/L_{\rm Edd}$ comes from the anecdotal evidence described by
Brandt, Pounds \& Fink (1995), Brandt et al. (1995), Grupe et al.
(1995), and Forster \& Halpern (1996) where a  
number of Seyfert galaxies with a steep $\alpha_x$ display
rapid, large amplitude, soft X-ray variability, which as Boller, Brandt,
\& Fink discuss may imply a low mass black hole, and thus a high 
$L/L_{\rm Edd}$. We are currently pursuing 
a more systematic study of the soft X-ray variability properties of broad
versus narrow line quasars using the {\em ROSAT} HRI.

The Pounds et al. suggestion is very appealing since it allows
a physical explanation for the tight correlation of apparently
completely unrelated quantities. Although it is not clear a priori
that $\alpha_x$ must steepen with increasing $L/L_{\rm Edd}$,
it appears that this is indeed what happens in Galactic black hole
candidates. 

Wang et al. also suggested that steep $\alpha_x$ objects have
a high $L/L_{\rm Edd}$ based on the fact that the fraction of luminosity
emitted in the X-ray regime in thin accretion disk models increases with
$L/L_{\rm Edd}$, as discussed above in \S 4.5. However, if this were
indeed the physical process behind the $\alpha_x$ versus \hb\ FWHM
correlation then one would expect high $L/L_{\rm Edd}$ objects to
have a high soft X-ray to optical flux ratio, while we find no correlation
between \hb\ FWHM and $\alpha_{os}$ ($r_S=-0.079$, Table 5).
 
We note in passing that one does not need to eliminate 
the normal broad line region in narrow line AGNs, as one of the
options suggested by
Boller et al., and Pounds et al. The lines are narrow simply 
because of the lower black hole mass. The broad line emitting gas
does not extend much closer to the center in narrow line AGNs,
as it does not extend much closer to the center in other AGNs, 
simply because of the effects of a higher
ionization parameter, and a higher gas density, each of which
quenches line emission.

\subsection {The X-ray Weak Quasars}

Two of the quasars in our sample, the RQQ PG~1001 +054 and PG~1411+442,
and possibly also the RLQ PG~1425+267 appear to form a distinct group which 
we term
here ``X-ray weak'' quasars, where the normalized X-ray luminosity is
a factor of 10-30 smaller than the sample median. The position of these
quasars as outliers can be noticed in the near IR normalized flux 
distribution (Fig.3), in the $\alpha_x$ versus $\alpha_{ox}$ correlation
(Fig.5e), and in the \hb\ versus 2~keV and 0.3~keV luminosity 
correlations (Fig.4). The first two indicators are based on the spectral 
shape, but the last one is independent of the spectral shape, and it also
suggests a deficiency of the X-ray luminosity by a factor of 10-30
relative to the one expected based on the \hb\ luminosity.
An apparently bimodal distribution in $\alpha_{ox}$ can also be seen
in Figure 5b of Wang et al. where 6 of their 86 quasars appear to form
a distinct group with $\alpha_{ox}<-2$.

No bimodality of $\alpha_{ox}$ is seen in the Avni \& Tananbaum (1986)
{\em EINSTEIN} study of the PG quasars. All the ``X-ray weak'' quasars 
found by the {\em ROSAT} PSPC have $\alpha_{ox}<-2$, but none of the quasars 
detectd by Tananbaum et al. have $\alpha_{ox}<-2$ (see Fig.8 in Avni \& 
Tananbaum). The lack of $\alpha_{ox}<-2$ and bimodality in the Tananbaum 
et al. sample probably reflects its incompleteness, as only 86\% of the 
quasars they observed were detected.

Although the three X-ray weak quasars in our sample 
stand out in luminosity correlations, they
conform well to the $\alpha_x$ correlations (Figs.5a-d). They thus have
the ``right'' slope but the ''wrong'' flux level. Why are these quasars
different?  A simple answer is that for some unknown reason the X-ray 
emission mechanism, most likely Comptonization by $T\ge 10^8$~K electrons,
tends to be bimodal, and in about 10\% of quasars 
(or in all quasars for $\sim 10$\% of the time) the X-ray flux
level is strongly suppressed, while the spectral slope is not affected.
Another option is that these are just normal quasars where the direct 
X-ray flux happens to be obscured. In this case what we see is only the
scattered X-ray flux. Photoionization calculations indicate that a few 
percent of the direct flux will be scattered, depending on the covering 
factor of the absorber and the ionization parameter. If the 
ionization parameter is large enough then the scattering will be 
mostly by free electrons which preserves the spectral shape 
(see Netzer 1993, and by Krolik \& Kriss 1995).
Such scattering will explain why the flux level is strongly reduced, while
the spectral shape is not affected. Note that the obscuring matter should
be transparent in the visible range, as is the case with the absorbing
matter in BALQSO.

Additional hints towards this 
interpretation come from the fact that PG~1411+442 is a broad absorption
line quasar (BALQSO, Malkan Green \& Hutchings, 1987), and the UV absorbing gas 
may also produce soft
X-ray absorption, as may also be the case in PHL~5200 
(Mathur Elvis \& Singh, 1995).
In addition, Green \& Mathur (1996) find that BALQSO observed by 
{\em ROSAT} have $\alpha_{ox}\lesssim -1.8$, i.e. as observed here in the 
X-ray weak RQQ.
Another
hint is provided by the fact that PG~1114+445 is also somewhat 
underluminous at 0.3~keV (Fig.3b), and this quasar is most likely seen
through a warm absorber. The X-ray weak quasars could therefore be more
extreme cases of PG~1114+445 and have an absorbing column which is large 
enough to completely absorb the direct soft X-ray emission. 

Note that PG~1425+267 is a RLQ, while all BALQSO are known to be RQQ 
(Stocke et al.
1992). It would thus seem implausible to suggest that PG~1425+267 is a BAL.
However, PG~1425+267 has about the same relatively steep $\alpha_{ox}$, 
compared to other RLQ, as in 3C~351 (Fiore et al. 1993),  
where X-ray absorption by warm gas is
observed together with resonance UV absorption lines (Mathur et al. 1994)
which are narrower than
in `proper' BALQSO.

Forthcoming HST spectra of all 23 quasars in our sample will allow us to test if
there is a one to one correspondence between X-ray weakness and broad absorption
lines, i.e. if all X-ray weak quasars are BALQSO, and not just that 
all BALQSO are X-ray weak, as strongly suggested by the
Green \& Mathur, and the Green et al. (1996) results.

A simple test of whether these are truly ``X-ray weak quasars'', or just 
normal highly absorbed quasars, can be done by looking at their hard X-ray
emission. If the X-ray column is below $10^{24}$~cm$^{-2}$ then the
obscuring material would become transparent at $E<10$~keV, and the 
observed hard X-ray emission will rise steeply above the cutoff
energy,  as seen in various highly absorbed AGNs, such as Mkn~3 (Iwasawa et 
al. 1994), NGC~5506 (Nandra \& Pounds 1994),  
NGC~6552 (Fukazawa, et al. 1994; Reynolds
et al. 1994), and NGC~7582 (Schachter et al. 1996). One may also expect 
significant spectral features produced by the obscuring material
(e.g. Matt et al. 1996), depending on the ionization state of this material.
 Forthcoming ASCA observations of PG~1411+442 and
PG~1425+267 will allow us to test this scenario.

Another prediction is that the X-ray weak quasars should show lower
variability compared with other quasars of similar X-ray luminosity.
This is because: 1). they are intrinsically more X-ray luminous, and 
variability amplitude tends to drop with increasing luminosity 
(Barr \& Mushotzky 1986; also Fig.9 in 
Boller, Brandt \& Fink). 2). the scattering medium must be significantly 
larger than the X-ray source, and short time scale variability will be 
averaged out. If the X-ray weak quasars are just due to large amplitude
intrinsic variability of the soft X-ray emission, as seen in some steep
narrow line Seyfert 1 galaxies (\S 4.7), then one may expect the exact
opposite behavior, i.e. these quasars may become significantly brighter 
at soft X-rays at some stage in the future.

\section {SUMMARY}
 
We defined a complete sample of 23 optically selected quasars which 
includes all
the PG quasars at $z\le 0.400$, 
and \nhgal$< 1.9\times 10^{20}$~cm$^{-2}$. Pointed {\em ROSAT} PSPC 
observations were made for all quasars, yielding high S/N spectra
for most objects. The high quality of the {\em ROSAT} spectra allows one 
to determine the best fitting $\alpha_x$ with about an order of 
magnitude higher precision
compared with previously available X-ray spectra. In this paper we report
the observations of 13 quasars not described in Paper I, analyze the
correlation of the X-ray properties of the complete sample with other 
emission properties, determine the mean X-ray spectra of low z quasars,
discuss the possible origin of the $\alpha_x$ versus \hb\ FWHM 
correlation, the nature of X-ray weak quasars, and the physical origin
of the soft X-ray emission. Our major results are the following:

\begin{enumerate}
\item The spectra of 22 of the 23 quasars are consistent, to within
$\sim 10-30$\%, with a single power-law model over the rest frame range 
$0.2-2$~keV. 
There is no evidence for significant soft excess emission with 
respect to the best fit power-law. We place a limit of 
$\sim 5\times 10^{19}$~cm$^{-2}$ on the amount of 
excess foreground absorption by cold gas in most of our quasars. The 
limits are
$\sim 1\times 10^{19}$~cm$^{-2}$ in the two highest S/N spectra. 

\item Significant X-ray absorption by partially ionized gas (``warm absorber'')
in quasars is rather rare, occurring for $\lesssim 5$\% of the population,
which is in sharp contrast to lower luminosity AGNs, where significant 
absorption probably occurs for $\sim 50$\% of the population.

\item The average soft X-ray spectral slope for RQQ is 
$\langle\alpha_x\rangle=-1.72\pm 0.09$, and it agrees remarkably
well with an extrapolation of the mean 1050\AA-350\AA\ continuum 
recently determined by Zheng et al. (1996) for $z>0.33$ quasars.
For RLQ $\langle\alpha_x\rangle=-1.15\pm 0.16$, which suggests that
RLQ quasars are weaker than RQQ below 0.2~keV, as suggested also
by the Zheng et al. mean RLQ continuum. These results suggest that 
there is no steep soft component below 0.2~keV.

\item Extensive correlation analysis of the X-ray continuum 
emission parameters with optical emission line parameters indicates 
that the strongest correlation
is between $\alpha_x$, and the \hb\ FWHM. A possible explanation for
this remarkably strong correlation is a dependence of $\alpha_x$ on 
$L/L_{\rm Edd}$, as observed in Galactic black hole candidates.

\item There appears to be a distinct class of ``X-ray weak'' quasars,
which form $\sim 10$\% of the population,
where the X-ray emission is smaller by a factor of 10-30 than expected
based on their luminosity at other bands, and on their \hb\ luminosity.

\item Thin accretion disk models cannot reproduce the observed
0.2-2~keV spectral shape, and they also cannot reproduce the tight
correlation between the optical and soft X-ray emission.

\item The H~I/He~I ratio in the ISM at high Galactic latitudes must be
within 20\%, and possibly within 5\%, of the total H/He ratio.
\end{enumerate}

The main questions raised by this study are:
\begin{enumerate}

\item What is the true nature of X-ray quiet quasars? Are these quasars
indeed intrinsically X-ray weak, or are they just highly absorbed but 
otherwise normal quasars?

\item What physical mechanism is maintaining the strong correlation between
the optical-UV and the soft X-ray continuum emission, or equivalently, 
maintaining a very small dispersion in the maximum possible 
far UV cutoff temperature? 

\item What is the physical origin for the strong correlations
between $\alpha_x$, and $L_{\rm [O~III]}$, Fe~II/\hb, and the peak 
[O~III] to \hb\ flux ratio?

\item Is the soft X-ray emission indeed related to the presence of radio
emission, or is it just a spurious relation and the primary effect is related 
to the \hb\ line width?  Or, put differently,
do RLQ and RQQ of similar \hb\ FWHM have similar $\alpha_x$?
\end{enumerate}

Extensions of the {\em ROSAT} PSPC survey described in this paper to the
hard X-ray regime with ASCA and SAX, to the UV with HST, and soft X-ray
variability monitoring with the {\em ROSAT} HRI, which are currently being 
carried out, may provide answers to some of the questions raised above. 
These studies
will also allow us to: 1) Test if steep $\alpha_x$ quasars have
a steep 2-10~keV slope, as expected based on the Pounds et al. $L/L_{\rm Edd}$
interpretations. 2) Test if soft X-ray variability is indeed strongly tied 
to the \hb\ FWHM, as expected if the \hb\ FWHM is an indicator of 
$L/L_{\rm Edd}$. 3) Explore the relation of the UV line emission properties 
to the ionizing spectral shape. 

\acknowledgments
We thank Niel Brandt, Hagai Netzer, Bev Wills and an anonymous
referee for many useful comments and suggestions.
This work was supported in part by NASA grants NAG 5-2087, NAG 5-1618, 
NAG5-2496, NAG 5-30934, NAGW 2201 (LTSA), and NASA contract NAS 8-30751. 
A. L. acknowledges support by LTSA grant NAGW-2144.

\end{document}